\documentclass[fleqn,usenatbib]{mnras}

\usepackage{newtxtext,newtxmath}  
\usepackage[T1]{fontenc}

\DeclareRobustCommand{\VAN}[3]{#2}
\let\VANthebibliography\thebibliography
\def\thebibliography{\DeclareRobustCommand{\VAN}[3]{##3}\VANthebibliography}

\usepackage{graphicx}	% Including figure files
\usepackage{amsmath}	% Advanced maths commands
\usepackage{multicol}

\title[The NGC 5466 Tidal Stream]{Revisit NGC 5466 Tidal Stream with $Gaia$, SDSS/SEGUE and LAMOST}

\author[Y. Yang et al.]{
Yong Yang,$^{1,2}$
Jing-Kun Zhao,$^{1}$\thanks{E-mail: zjk@bao.ac.cn}
Miho N. Ishigaki,$^{3,4,5}$
Jian-Zhao Zhou,$^{6}$
Cheng-Qun Yang,$^{7}$
\newauthor{
Xiang-Xiang Xue,$^{1}$
Xian-Hao Ye$^{1,2}$
and Gang Zhao$^{1,2}$
}\\
$^{1}$CAS Key Laboratory of Optical Astronomy, National Astronomical Observatories, Chinese Academy of Sciences, Beijing 100101, People's Republic of China\\
$^{2}$School of Astronomy and Space Science, University of Chinese Academy of Sciences, Beijing 100049, People's Republic of China\\
$^{3}$National Astronomical Observatory of Japan, 2-21-1 Osawa, Mitaka, Tokyo 181-8588, Japan\\
$^{4}$Astronomical Institute, Tohoku University, 6-3, Aramaki, Aoba-ku, Sendai, Miyagi 980-8578, Japan\\
$^{5}$Kavli Institute for the Physics and Mathematics of the Universe (WPI), The University of Tokyo Institutes for Advanced Study, The University of Tokyo, 5-1-5\\
Kashiwanoha, Kashiwa, Chiba 277-8583, Japan\\
$^{6}$Department of Astronomy, Beijing Normal University, Beijing 100875, People's Republic of China\\
$^{7}$Key Laboratory for Research in Galaxies and Cosmology, Shanghai Astronomical Observatory, 80 Nandan Road, Shanghai 200030, People's Republic of China
}

\date{Accepted XXX. Received YYY; in original form ZZZ}

\pubyear{2021}

\begin{document}
\label{firstpage}
\pagerange{\pageref{firstpage}--\pageref{lastpage}}
\maketitle

\begin{abstract}
By mining the data from $Gaia$ EDR3, SDSS/SEGUE DR16 and LAMOST DR8, 11 member stars of the NGC 5466 tidal stream are detected and 7 of them are newly identified. To reject contaminators, a variety of cuts are applied in sky position, color-magnitude diagram, metallicity, proper motion and radial velocity. We compare our data to a mock stream generated by modeling the cluster's disruption under a smooth Galactic potential plus the Large Magellanic Cloud (LMC). The concordant trends in phase-space between the model and observations imply that the stream might have been perturbed by LMC. The two most distant stars among 11 detected members trace the stream's length to $60\degr$ of sky, supporting and extending the previous length of $45\degr$. Given that NGC 5466 is so distant and potentially has a longer tail than previously thought, we expect that NGC 5466 tidal stream could be a useful tool in constraining the Milky Way gravitational field. 
\end{abstract}


%%%%%%%%%%%%%%%%%%%%%%%%%%%%%%%%%%%%%%%%%%%%%%%%%%

%%%%%%%%%%%%%%%%% BODY OF PAPER %%%%%%%%%%%%%%%%%%

\section{Introduction}
\label{sec:introduction}

Stellar streams have been identified as the relics disrupted from globular clusters or dwarf galaxies \citep[e.g.,][]{2001ApJ...547L.133I,2001ApJ...548L.165O,2002ApJ...569..245N,2004AJ....128..245M,2009ApJ...693.1118G,2021ApJ...920...51M}. The discoveries of stellar streams offer an opportunity to probe the shape of the Galactic gravitational field \citep[e.g.,][]{2015ApJ...803...80K,2019MNRAS.486.2995M} and the formation history of the stellar halo \citep[e.g.,][]{2005ApJ...635..931B,2008ApJ...680..295B,2010ApJ...712..260K,2010ApJ...714..229L,2015MNRAS.449.1391B,2016ASSL..420...87G,2016MNRAS.463.1759B,2018MNRAS.481.3442M}. In addition, the scenario of hierarchical formation for galaxies in the standard $\Lambda$CDM cosmological model is supported by the existence of stellar streams \citep[e.g.,][]{2008MNRAS.391.1685S,2022ApJ...926..107M}. 

Streams formed from globular clusters are typically thin and dynamically cold, and have proved to be a powerful tool for studying the Milky Way potential \citep[e.g.,][]{2013MNRAS.436.2386L,2015ApJ...803...80K,2016ApJ...833...31B,2019MNRAS.486.2995M}. An example is the NGC 5466 tidal stream, which was independently detected by \citet{2006ApJ...637L..29B} and \citet{2006ApJ...639L..17G}. In \citet{2006ApJ...637L..29B}, a $\sim 4\degr$ stream was reported using neural networks to extract the probability distribution of the cluster stars. However, by applying an optimal matched filter technique to SDSS photometric data, \citet{2006ApJ...639L..17G} found a $45\degr$ stream extending from $\alpha \simeq 180\degr$ to $\alpha \simeq 225\degr$. The authors also noted that a part of stream with $\alpha < 190\degr$ did not lie perfectly along computed orbits of the globular cluster but deviated from a smooth curve. They attributed this ``deviation'' to irregularities of the Galactic halo potential or a weak encounter with a mass concentration in the disk. The claim of $45\degr$ stream received some support from the results modelled by \citet{2007MNRAS.380..749F}, in which $\ga 100\degr$ tails was produced. By investigating the ``deviation'', \citet{2012MNRAS.424L..16L} examined the constrains provided by the stream on the Galactic halo shape and pointed out that only for either oblate or fully triaxial halo potentials such ``deviation'' was possible. 

Although a long stellar stream around NGC 5466 was reported, no individual members of the stream have been distinguished until two recent pioneering works. In \citet{2021ApJ...914..123I}, more than 20 member stars of the stream were successfully detected for the first time with \texttt{STREAMFINDER} algorithm \citep{2018MNRAS.477.4063M}. In \citet{2021MNRAS.507.1923J}, 33 member stars across $\sim 31\degr$ of sky were identified through position, proper motion and metallicity cuts. The later work also compared the observed samples in phase-space to the mock stream generated in a simplified axisymmetric potential model, showing concordance between them. However, they considered that the signals at $\alpha \la 200\degr$ in \citet{2006ApJ...639L..17G} might be not real, where few stream members were detected and the trajectory of their model deviated from the claimed matched filter path. 

Given the contending conclusions about the stream's length and its significant functionality in unraveling the Milky Way halo shape \citep{2012MNRAS.424L..16L,2013MNRAS.436.2386L}, in this work we aim to search for supplementary member stars, especially for those located far from the globular cluster, to provide more conclusive results on the length and trajectory of the NGC 5466 tidal stream. The paper is organized as follows. In Section~\ref{sec:data}, we introduce the data. Section~\ref{sec:search} describes the search strategy for member stars of the stream. The model of the cluster's disruption is given in Section~\ref{sec:mockstream}. We present a discussion in Section~\ref{sec:discussion} and draw our conclusion in Section~\ref{sec:summary}.

\section{Data}
\label{sec:data}

We base our search on high-quality astrometry and photometry provided by the $Gaia$ EDR3 \citep{2021A&A...649A...1G,2021A&A...649A...4L,2021A&A...649A...3R}, along with the spectroscopic data from the Sloan Extension for Galactic Understanding and Exploration \citep[SEGUE;][]{2009AJ....137.4377Y} and the Large Sky Area Multi-Object Fiber Spectroscopic Telescope \citep[LAMOST;][]{2012RAA....12.1197C,2006ChJAA...6..265Z,2012RAA....12..723Z,2015RAA....15.1089L} surveys.

We begin by retrieving all stars satisfying $180\degr < \alpha < 225\degr$ and $18\degr < \delta < 45\degr$ from the $Gaia$ EDR3 catalogue as illustrated in Fig.~\ref{fig:alpha_delta}, considering the known extent of the NGC 5466 tidal stream on the celestial sphere \citep{2016ASSL..420.....N}. The zero-point correction in the parallax is implemented using the code provided by \citet{2021A&A...649A...4L}. The corrections of the G-band magnitude and the photometric flux excess factor are applied as instructed in \citet{2021A&A...649A...3R}. In order to ensure good astrometric and photometric solutions, only stars with \texttt{ruwe} < 1.4 and corrected BP/RP excess factor smaller than 3 times the associated uncertainty \citep[see Section 9.4 in][]{2021A&A...649A...3R} are retained.

Given that the distance of NGC 5466 cluster is about 16 kpc \citep[2010 edition]{1996AJ....112.1487H}, we remove all of the foreground stars within a distance of 10 kpc from the Sun, with 3 sigma confidence interval with the selection $\varpi - 3\sigma_{\varpi} > 0.1$ mas. Stars that do not meet this criteria, including ones without $\varpi$ or with negative $\varpi$, are still retained since we are not sure enough whether they are within 10 kpc. The remaining stars are cross-matched with SDSS/SEGUE DR16 \citep{2012ApJS..203...21A}\footnote{SEGUE data have not been updated since SDSS DR9 but are included in subsequent data releases.} and LAMOST DR8, by which the metallicity and heliocentric radial velocity are obtained.

\section{Stream Member Stars Selection}
\label{sec:search}

In this section, we try to identify the member stars of the NGC 5466 tidal stream using the data yielded from the Section~\ref{sec:data}.

\subsection{Sky Position}
\label{subsec:sky}

\begin{figure*}
	\includegraphics[width=0.8\linewidth]{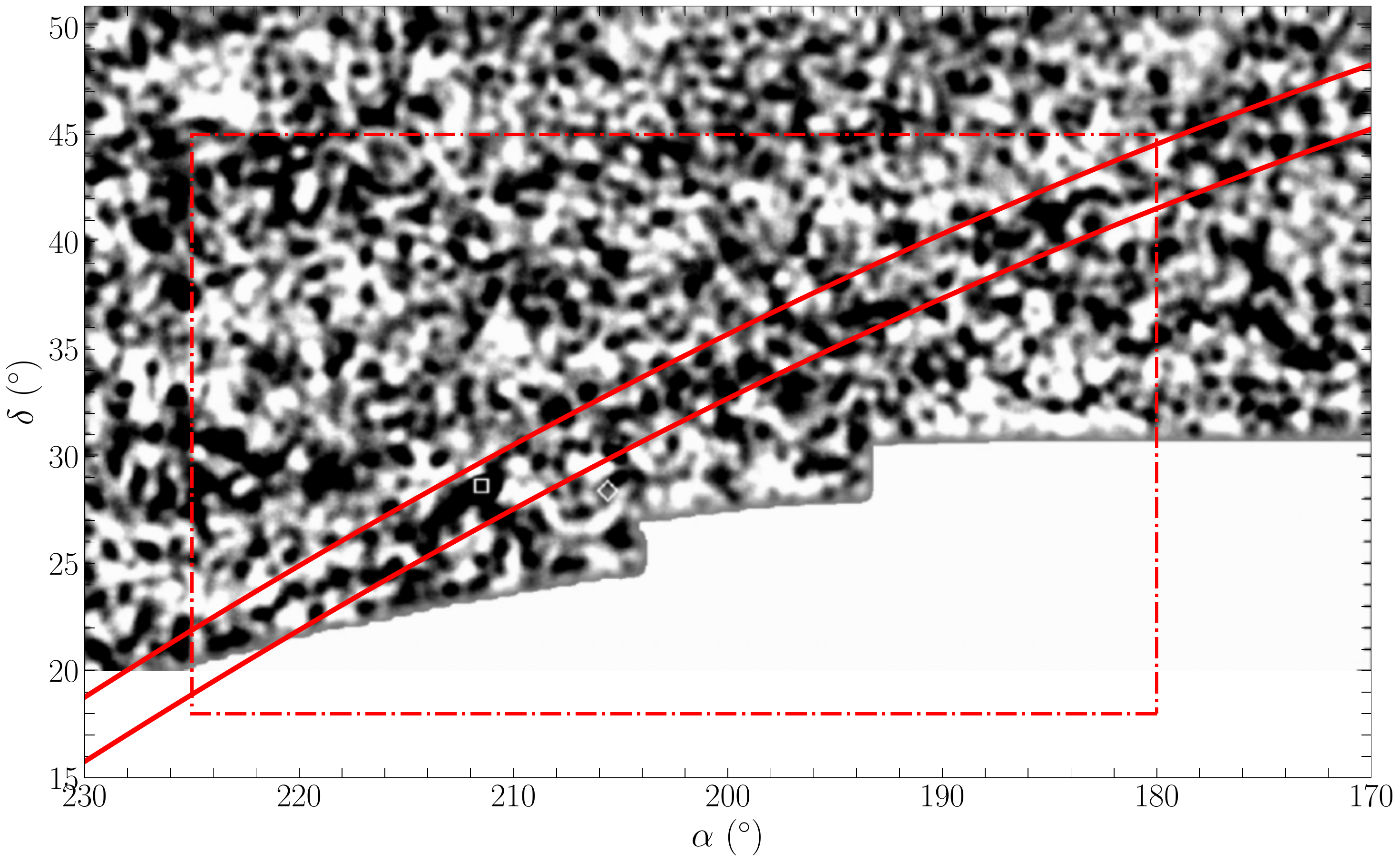}
    \caption{A comparison between the $45\degr$ detection in \citet{2006ApJ...639L..17G} and the region used to constrain the sky position of stars. The red solid lines correspond to equation~(\ref{eq:fit}) $\pm 1.5\degr$. The red rectangle corresponds to the sky extent of the data retrieved in Section~\ref{sec:data}. The open square and diamond represent the location of NGC5466 and NGC 5272 cluster.}
    \label{fig:alpha_delta}
\end{figure*}

Although we have retrieved the stars inside a rectangle corresponding to the known extent of the stream, there are still too many field stars that need to be cleaned up. We achieve this by fitting the track of the stream with a second-order polynomial and retaining the stars in the region created by moving the polynomial, which is given by equation~(\ref{eq:fit}), up and down by $1.5\degr$ along the $\delta$ direction. It can be seen in Fig.~\ref{fig:alpha_delta} that the the stream trajectory is well enclosed, including the ``deviation'' part at $\alpha \simeq 185\degr$. Through this selection, 3,130 stars inside the region are left.

\begin{equation}
    \delta = -0.0024243\alpha^2 + 0.47813253\alpha + 35.54016132
    \label{eq:fit}
\end{equation}

\subsection{Color-Magnitude Diagram}

\begin{figure}
	\includegraphics[width=\columnwidth]{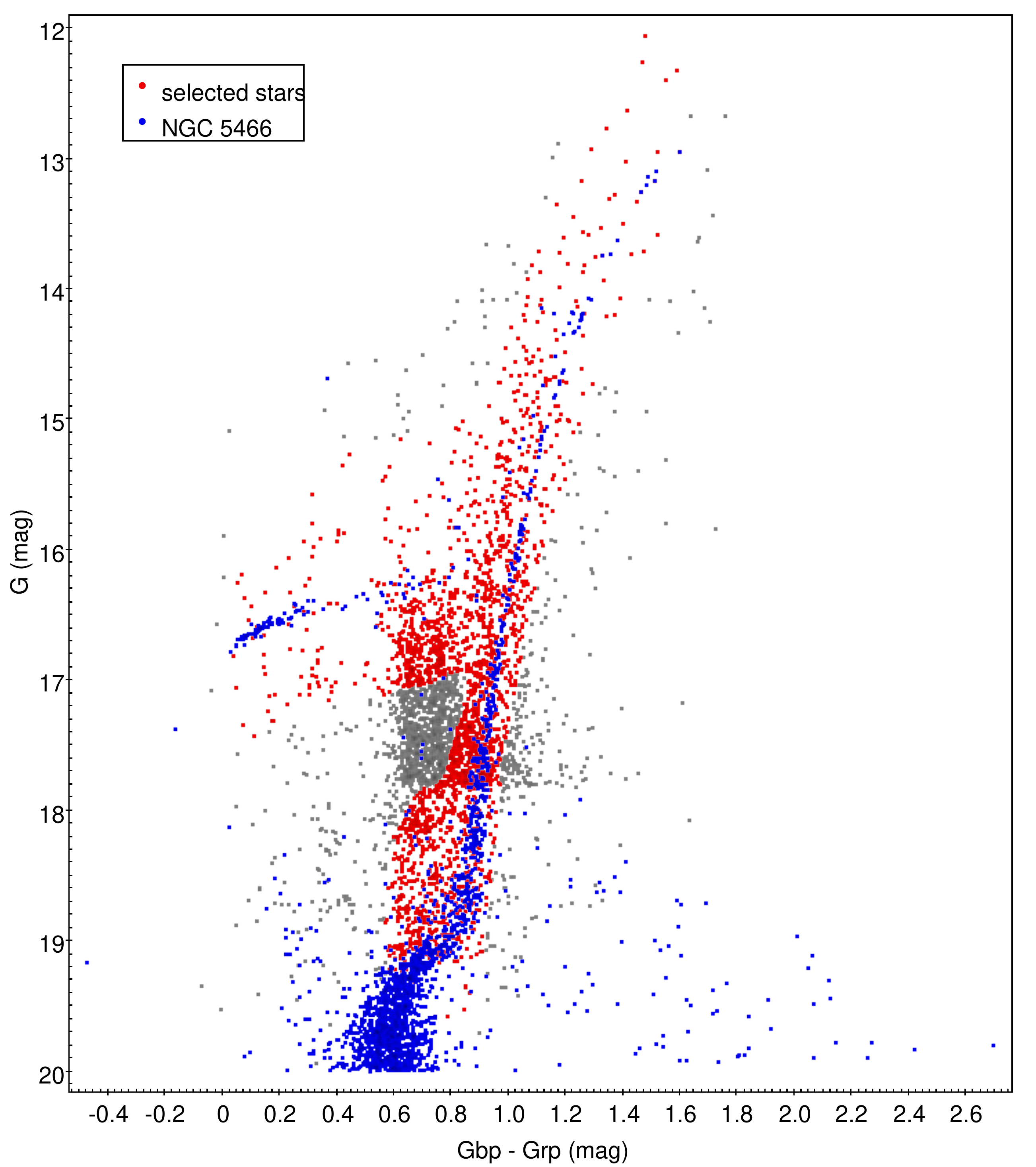}
    \caption{The grey dots represent the stars that pass the first selection. The blue dots represent the stars within the tidal radius of NGC 5466. The selected stars during this step are plotted with the red dots.}
    \label{fig:CMD}
\end{figure}

We then select candidates of member stars in color-magnitude diagram (CMD), as illustrated in Fig.~\ref{fig:CMD}. All sources here have been extinction-corrected using the \citet{1998ApJ...500..525S} maps as re-calibrated by \citet{2011ApJ...737..103S} with RV = 3.1, assuming $A_{G}/A_{V} = 0.83627$, $A_{BP}/A_{V} = 1.08337$, $A_{RP}/A_{V} = 0.63439$\footnote{These extinction ratios are listed on the Padova model site \url{http://stev.oapd.inaf.it/cgi-bin/cmd}}. The grey dots represent the stars that pass the first selection. The blue dots are those stars within the tidal radius of NGC 5466 $r_t = 15.68'$ \citep[computed from][2010 edition]{1996AJ....112.1487H}. We use them as the ``isochrone'' of the globular cluster. Considering a possible wide range of distances traced by the stream, we shift the ``isochrone'' up/down by 1.02 / 0.69 mag, corresponding to moving the cluster to a heliocentric distance of 10 / 22 kpc, which is also the extent inspected by \citet{2021MNRAS.507.1923J}. We pick out the stars within the area swept by the ``isochrone'' and show them with the red dots in Fig.~\ref{fig:CMD}. This selection yields 1,960 stars.

\subsection{Metallicity}

Given that NGC 5466 has a [Fe/H] of -1.98 dex \citep[2010 edition]{1996AJ....112.1487H}, we apply a conservative metallicity cut of -2.5 dex < [Fe/H] < -1.5 dex to the remaining candidates, among which there is a maximum of [Fe/H] uncertainty of 0.5 dex. Through this cut, 570 stars are remained.

\subsection{Radial Velocity}

Fitting the radial velocity distribution to single out the stream stars has been used in previous studies \citep[e.g.,][]{2010ApJ...712..260K}. Here we mainly refer to the Bayesian method in \citet{2019ApJ...877...13H}.

\begin{figure}
	\includegraphics[width=\columnwidth]{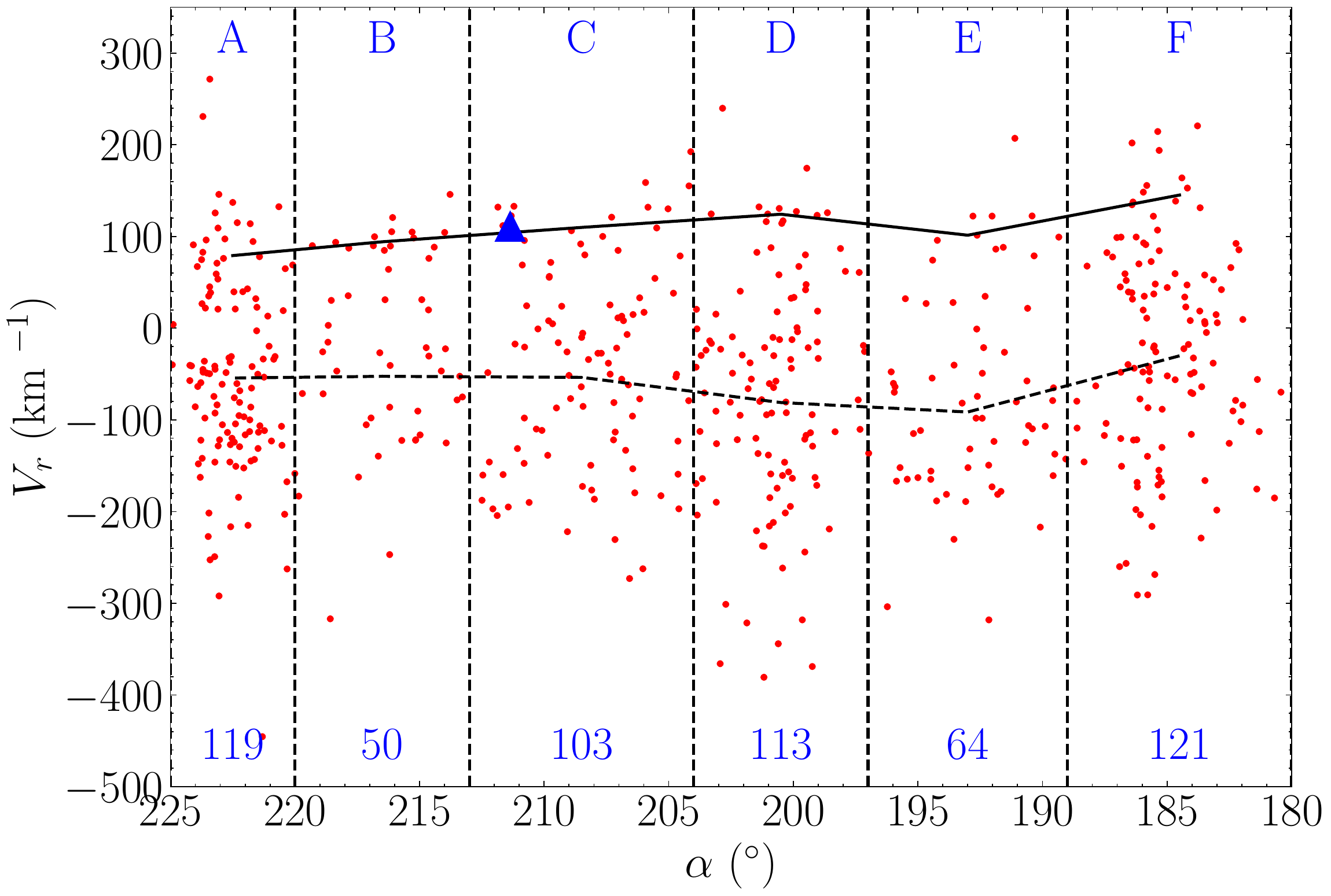}
    \caption{The diagram of $\alpha - V_r$ of the remaining stars, which are divided into six regions in $\alpha$ from A to F. The number of stars in each $\alpha$ bin is marked on the bottom. The location of NGC 5466 cluster is marked with a blue triangle. The solid and dashed lines show the trends of $V_r^{st}$ and $V_r^{mw}$ based on the Bayesian results from Table~\ref{tab:results}}
    \label{fig:alpha_Vr}
\end{figure}

We firstly divide the stars in $\alpha - V_r$ plane into six regions, as shown in Fig.~\ref{fig:alpha_Vr}. These divisions are just based on the visual differences in number density of the stars: higher density for A, D and F, lower density for B and E, and moderate density for region C. Unlike \citet{2019ApJ...877...13H}, we directly use the heliocentric radial velocities instead of converting them to the Galactic frame. For each region, it is assumed that the velocity distribution can be described as a combination of two Gaussians, one for the stream stars and the other for the Milky Way field stars. The component of stream stars has a mean $V_r^{st}$, a dispersion $\sigma_{V_r}^{st}$ and a fraction $f^{st}$, while the component of field stars has a mean $V_r^{mw}$ and a dispersion $\sigma_{V_r}^{mw}$. With the model parameters $\Theta = \{f^{st},V_r^{st},\sigma_{V_r}^{st},V_r^{mw},\sigma_{V_r}^{mw}\}$, the likelihood of observing a star with a radial velocity of $V_r^i$ is given by

\begin{equation}
    \mathcal{L}^i(V_r^i|\Theta) = f^{st}\mathcal{N}(V_r^i|V_r^{st},\sigma_{V_r}^{st}) + (1-f^{st})\mathcal{N}(V_r^i|V_r^{mw},\sigma_{V_r}^{mw}),
    \label{eq:likelihood}
\end{equation}
where $\mathcal{N}(x|\mu,\sigma)$ denotes a Gaussian function: 

\begin{equation}
    \mathcal{N}(x|\mu,\sigma) = \frac{1}{\sqrt{2\pi}\sigma}{\rm exp}\left[ -\frac{(x - \mu)^2}{2\sigma^2} \right].
\end{equation}
By multiplying the likelihood of each star in a certain region, we can obtain the total likelihood of this region $\mathcal{L}(V_r|\Theta)$. The posterior distribution is given by

\begin{equation}
    \mathcal{P}(\Theta|V_r) \varpropto \mathcal{L}(V_r|\Theta) \mathcal{I}(\Theta),
    \label{eq:posterior}
\end{equation}
where $\mathcal{I}(\Theta)$ represents the priors of the model parameters. A uniform distribution is adopted for the priors of all parameters of each region and details of them are listed in Table~\ref{tab:prior}.

\begin{table}
	\centering
	\caption{Priors of the model parameters of each region.}
	\label{tab:prior}
	\begin{tabular}{cccccc} %  alignment for each
		\hline
		Region & $f^{st}$ & $V_r^{st}$ & $\sigma_{V_r}^{st}$ & $V_r^{mw}$ & $\sigma_{V_r}^{mw}$ \\
		       &        & (km\,s$^{-1}$) & (km\,s$^{-1}$) & (km\,s$^{-1}$) & (km\,s$^{-1}$) \\
		\hline
		A & [0, 1] & [0, 150] & [0, 20] & [-150, 150] & [50, 200]\\
		B & [0, 1] & [0, 150] & [0, 20] & [-150, 150] & [50, 200]\\
		C & [0, 1] & [0, 150] & [0, 20] & [-150, 150] & [50, 200]\\
		D & [0, 1] & [50, 150] & [0, 20] & [-150, 150] & [50, 200]\\
		E & [0, 1] & [50, 150] & [0, 20] & [-150, 150] & [50, 200]\\
		F & [0, 1] & [120, 180] & [0, 20] & [-150, 150] & [50, 200]\\
		\hline
	\end{tabular}
\end{table}

\begin{figure*}
	\includegraphics[width=\linewidth]{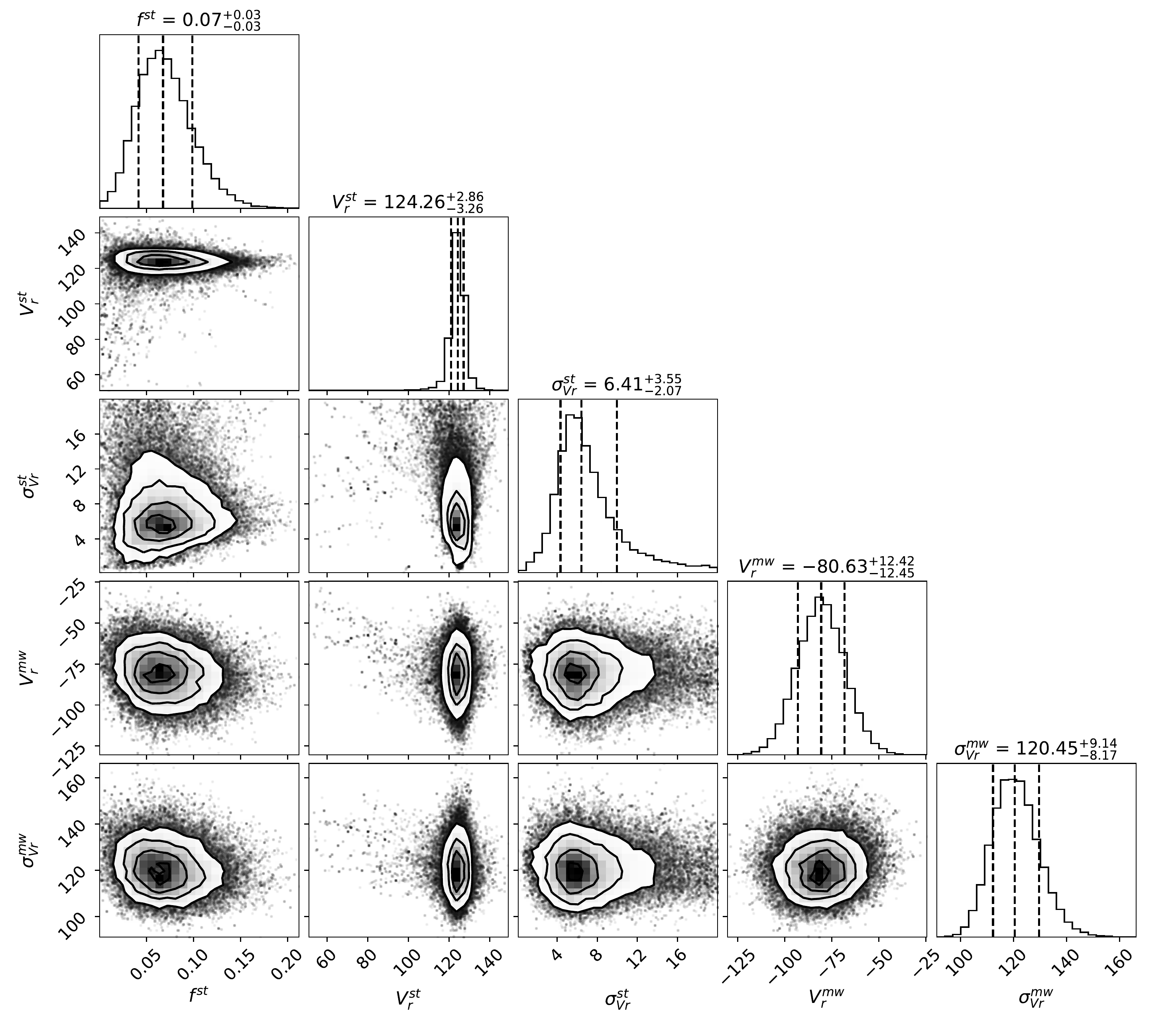}
    \caption{The marginalized posterior distributions of the model parameters for region D obtained with MCMC. Three vertical dashed lines correspond to the 16th, 50th and 84th percentiles of each posterior. The unit of velocity and its dispersion is km\,s$^{-1}$.}
    \label{fig:mcmc}
\end{figure*}

In order to obtain marginalized posterior distributions of the model parameters, we run the Markov chain Monte Carlo (MCMC) code \citep{2013PASP..125..306F} for each region. As an example, Fig.~\ref{fig:mcmc} shows the resulting distributions for region D, where the three vertical dashed lines indicate the 16th, 50th and 84th percentiles of the posteriors. The medians of all posteriors are summarized in Table~\ref{tab:results}.

\begin{table}
	\centering
	\caption{The medians of model parameters of each region.}
	\label{tab:results}
	\begin{tabular}{cccccc} %  alignment for each
		\hline
		Region & $f^{st}$ & $V_r^{st}$ & $\sigma_{V_r}^{st}$ & $V_r^{mw}$ & $\sigma_{V_r}^{mw}$ \\
		       &        & (km\,s$^{-1}$) & (km\,s$^{-1}$) & (km\,s$^{-1}$) & (km\,s$^{-1}$) \\
		\hline
		A & 0.03 & 79.16 & 12.73 & -54.33 & 110.37\\
		B & 0.23 & 94.21 & 10.62 & -52.51 & 96.45\\
		C & 0.17 & 109.88 & 13.81 & -53.64 & 105.99\\
		D & 0.07 & 124.26 & 6.41 & -80.63 & 120.45\\
		E* & 0.10 & 101.38 & 14.86 & -91.44 & 101.04\\
		F & 0.02 & 145.18 & 11.54 & -30.15 & 116.13\\
		\hline
	\end{tabular}
\end{table}

For region A, B and C, a general prior [0, 150] km\,s$^{-1}$ is adopted for $V_r^{st}$ and a monotonous trend that $V_r^{st}$ increases when $\alpha$ decreases is obtained after running MCMC. The other priors of $V_r^{st}$ especially for region F are aimed to ensure this trend, which is crucial for estimating stream’s parameters. Generally, the radial velocities of a halo stream are supposed to change monotonically along coordinates as long as there is no turning point contained (like apogalacticon), such as Pal 5 \citep{2016ApJ...823..157I}, GD-1 \citep{2016ApJ...833...31B}, Hr\'{i}d and Gj\"{o}ll stream \citep{2021ApJ...914..123I}. Therefore, an assumption of a monotonous trend in $\alpha - V_r$ plane might be more reasonable for NGC 5466 stream than a non-monotonic one and it is also supported by the later model of the cluster's disruption (see Section~\ref{sec:mockstream}). 

In Fig.~\ref{fig:alpha_Vr}, we show the trends of $V_r^{st}$ and $V_r^{mw}$ with the solid and dashed lines based on the results from Table~\ref{tab:results}. We note that region E has an incompatible $V_r^{st}$ = 101.38 km\,s$^{-1}$ which should be between that of region D and F [124.26, 145.18] km\,s$^{-1}$, given the trend assumption above. Furthermore, there are no stars at all in region E located within this $V_r^{st}$ range according to Fig.~\ref{fig:alpha_Vr}. Hence, we consider that no stream members in this region are detected in our samples and the corresponding parameters derived from MCMC do not represent real situation of the stream. Instead, we fit them with one Gaussian giving a mean $V_r^{mw}$of -72.85 km\,s$^{-1}$ and a dispersion $\sigma_{V_r}^{mw}$ of 109.63 km\,s$^{-1}$. 

\begin{figure*}
	\includegraphics[width=\linewidth]{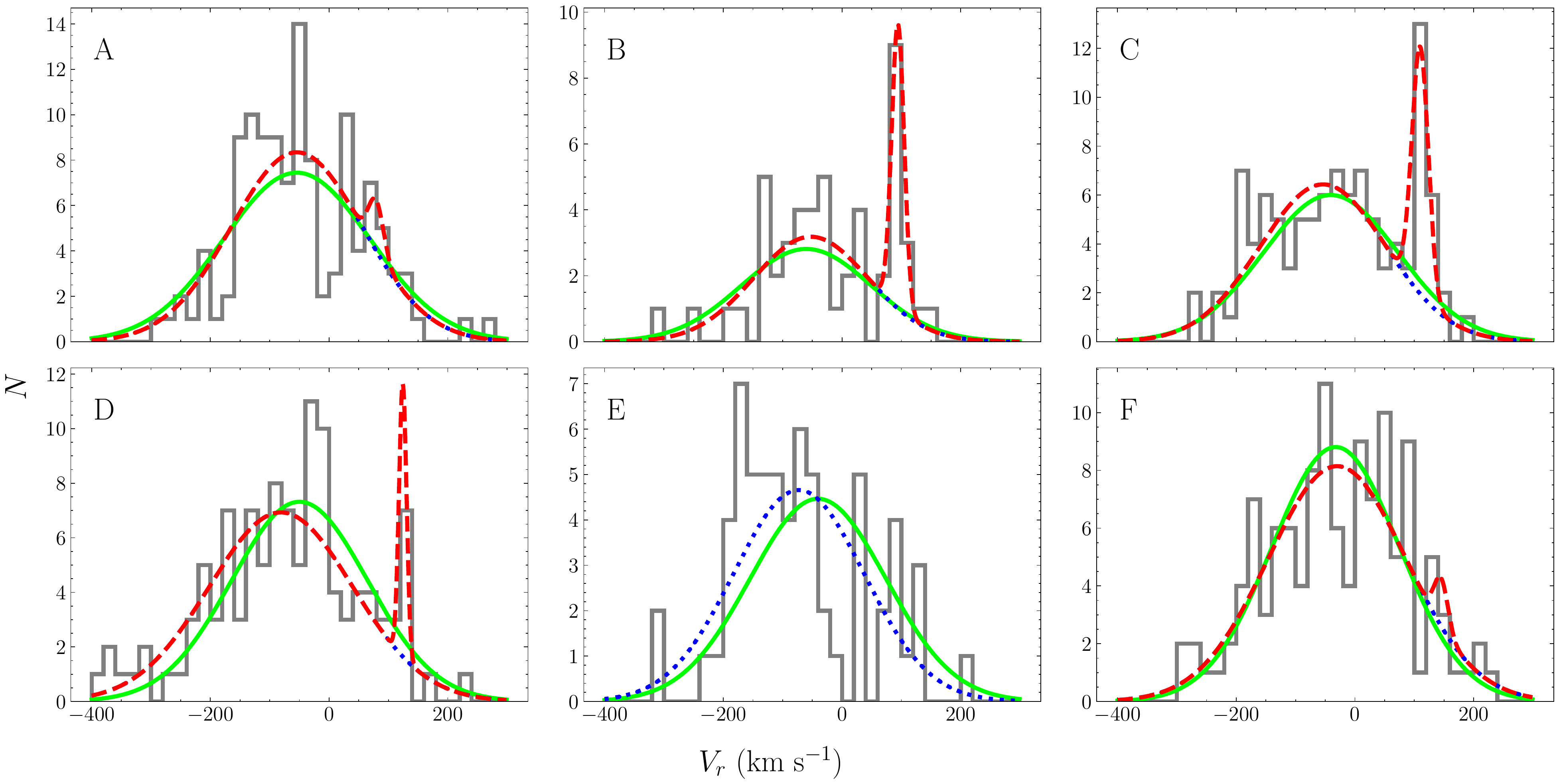}
    \caption{The histograms of $V_r$ of region A to F. The over plotted lines represent the fitted distributions: the stream and the Milky Way field stars (red dashed), only the Milky Way field stars (blue dotted), and the simulated field stars (green solid). For region E, the first distribution is not plotted since no stream stars are detected in this region.}
    \label{fig:hists}
\end{figure*}

To validate the derived parameters, synthetic stellar population similar to the data are generated using the Besan\c{c}on model\footnote{\url{https://model.obs-besancon.fr}} \citep{2003A&A...409..523R} and fitted with a Gaussian function as well. Comparisons between observed and synthetic data are shown in Fig.~\ref{fig:hists}, in which the distributions of field stars are generally consistent with the simulated ones.

With the model parameters in Table~\ref{tab:results}, we further select the member stars whose radial velocities satisfy $|V_r - V_r^{st}| < 3\sigma_{V_r}^{st}$, with which 89 stars are retained.

\subsection{Proper Motion}

\begin{figure*}
    \includegraphics[width=0.9\linewidth]{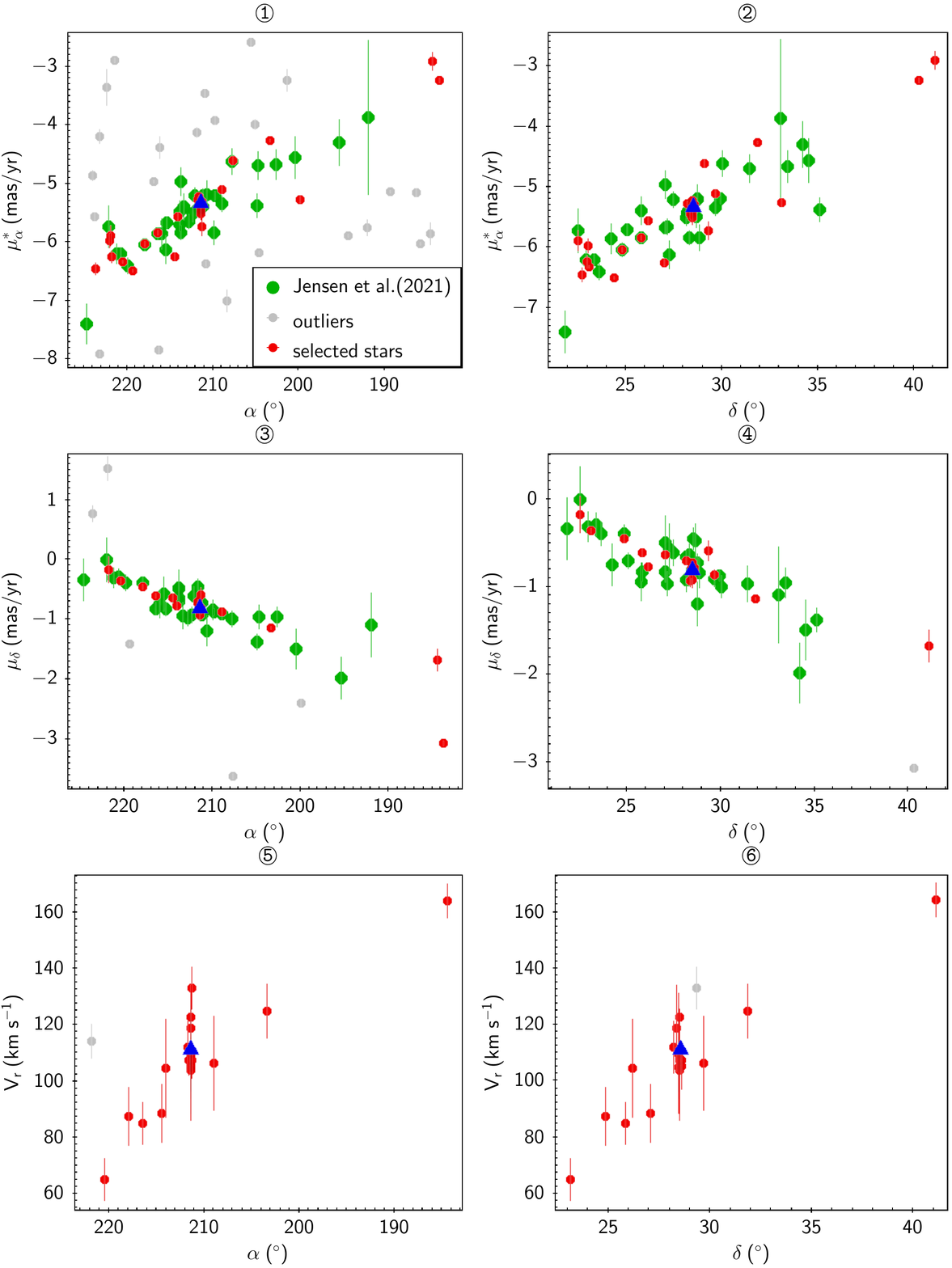}
    \caption{The proper motion in $\alpha$ and $\delta$, and radial velocity as a function of $\alpha$ and $\delta$, are shown from the top to the bottom, respectively. The red points represent those selected stars consistent with the members from \citet{2021MNRAS.507.1923J} marked in green points. The gray points represent rejected outliers. The blue triangle represents NGC 5466 cluster, whose proper motions are taken from \citet{2021MNRAS.505.5978V}.}
    \label{fig:relation}
\end{figure*}

As a stellar stream tidally disrupted from NGC 5466, it is expected to show a coherent sequence both in space and kinematics. The rest of contaminators are further rejected based on this, which is illustrated in Fig.~\ref{fig:relation}. The member stars (green) found by \citet{2021MNRAS.507.1923J} are employed as an indicator of the stream sequence in \{$\alpha$, $\delta$\} vs. \{$\mu_{\alpha}^*$, $\mu_{\delta}$\} planes. The outliers (gray) that deviate from the stream sequence are removed. Those stars (red) consistent with the sequence are retained and proceed to be investigated in the next plane. This means that the red plus gray points in plane $N$ come from the red points in plane $N$-1. $V_r$ is further used to exclude contaminators. In such a step-by-step manner, 19 highly probable member stars of NGC 5466 are finally obtained, 9 out of which are outside the tidal radius $r_t$. 

\section{Comparing to Mock Streams}
\label{sec:mockstream}

In order to evaluate reliability of the selected stars, we consider a model of the NGC 5466 cluster's disruption to inspect the differences between observation and theory. This is achieved with the Python package \texttt{Gala} \citep{2017JOSS....2..388P} which is designed for performing common tasks needed in Galactic Dynamics.

\subsection{A Smooth Milky Way Potential}
\label{subsec:smooth}

A smooth Galactic potential model is firstly examined, where ``smooth'' means that the sub-halo is not considered. The model consists of a Plummer bulge \citep{1911MNRAS..71..460P}, $\Phi_{\rm bulge}$, two Miyamoto-Nagai disks \citep{1975PASJ...27..533M}, $\Phi_{\rm thin}$ and $\Phi_{\rm thick}$, and a spherical NFW halo \citep{1996ApJ...462..563N}, $\Phi_{\rm halo}$: 

\begin{equation}
    \Phi_{\rm bulge}(r) = \frac{-GM_{\rm bulge}}{\sqrt{r^2+b_{\rm bulge}^2}}
\end{equation}

\begin{equation}
    \Phi_{\rm thin}(R,z) = \frac{-GM_{\rm thin}}{\sqrt{R^2+(a_{\rm thin}+\sqrt{z^2+b_{\rm thin}^2})^2}}
\end{equation}

\begin{equation}
    \Phi_{\rm thick}(R,z) = \frac{-GM_{\rm thick}}{\sqrt{R^2+(a_{\rm thick}+\sqrt{z^2+b_{\rm thick}^2})^2}}
\end{equation}

\begin{equation}
    \Phi_{\rm halo}(r) = \frac{-4\pi G \rho_s r_s^3}{r} {\rm ln}(1+\frac{r}{r_s})
\end{equation}
where $r$ is the Galactocentric radius, $R$ is the cylindrical radius and $z$ is the vertical height. For the bulge and disks, we adopt the parameters from \citet[][Model I]{2017A&A...598A..66P}. The virial mass $M_{\rm virial}$ and concentration $c$ used to initialize the NFW halo are from \citet{2017MNRAS.465...76M}. Those chosen parameters are summarized in Table~\ref{tab:potential}. 

\begin{table}
	\centering
	\caption{Adopted parameters of the Galactic potential.}
	\label{tab:potential}
	\begin{tabular}{cc}
		\hline
		Parameter & Value \\
		\hline
		$M_{\rm bulge}$ &  $1.0672\times 10^{10} M_{\sun}$  \\
		$b_{\rm bulge}$ &  0.3 kpc  \\
		$M_{\rm thin}$ &  $3.944\times 10^{10} M_{\sun}$  \\
		$a_{\rm thin}$ &  5.3 kpc  \\
		$b_{\rm thin}$ &  0.25 kpc  \\
		$M_{\rm thick}$ &  $3.944\times 10^{10} M_{\sun}$  \\
		$a_{\rm thick}$ &  2.6 kpc  \\
		$b_{\rm thick}$ &  0.8 kpc  \\
		$M_{\rm virial}$ &  $1.37\times 10^{12} M_{\sun}$  \\
		$c$  &  15.4  \\
		\hline
	\end{tabular}
\end{table}

As for the internal gravity of the globular cluster NGC 5466, we choose a Plummer potential: 

\begin{equation}
    \Phi_{\rm GC}(r') = \frac{-GM_{\rm GC}}{\sqrt{r'^2+b_{\rm GC}^2}}
\end{equation}
with a $M_{\rm GC}$ of $6\times10^4 M_{\sun}$ \citep{2017MNRAS.471.3668S} and a $b_{\rm GC}$ of 6.7 pc computed from the core radius of the cluster \citep[2010 edition]{1996AJ....112.1487H}. Here $r'$ denotes the distance to the cluster's center. 

The solar distance to the Galactic center, circular velocity at the Sun and solar velocities relative to the Local Standard of Rest are set to 8 kpc, 220 km\,s$^{-1}$ \citep{2012ApJ...759..131B} and (11.1, 12.24, 7.25) km\,s$^{-1}$ \citep{2010MNRAS.403.1829S}, respectively. To generate long enough tidal tails which can completely cover the data, the cluster is initialized 5 Gyr ago and integrated forward from then on, releasing two particles (leading and trailing directions respectively) at Lagrange points \citep{2014MNRAS.445.3788G} per 1 Myr with a total of 5000 steps. The velocity dispersion is set to 1.6 km\,s$^{-1}$ \citep{2018MNRAS.478.1520B} and the cluster mass is fixed during this process. By doing so, a mock NGC 5466 stream is obtained and we then compare it to observed data in phase-space. 

In the left panel of Fig.~\ref{fig:comparison_smooth_LMC}, the mock stream, the member stars found by the above selections and \citet{2021MNRAS.507.1923J}, are shown in the orange dots, red squares and green points, respectively. The black crosses represent two contaminators that are further discerned given their heliocentric distance and metallicity (see Table~\ref{tab:members}). Additional members marked with the red diamonds are further selected based on the mock stream which will be described in Section~\ref{subsec:LMC}. Those 10 stars inside the tidal radius $r_t = 15.68'$, which are still bound to the cluster, are not shown here. The NGC 5466 cluster is still marked in a blue triangle. The distances can be obtained from \citet{2019A&A...628A..94A} except for the star at $\alpha \simeq 184.40\degr$ (we call it ``S1'' for convenience). We estimate a distance to S1 of $17.37^{+5.74}_{-3.45}$ kpc using the method from \citet{2015AJ....150....4C}, which is a Bayesian approach with likelihood estimated via comparison of spectroscopically derived atmospheric parameters to a grid of stellar isochrones, and returns a posterior probability density function for star's absolute magnitude. 

From Fig.~\ref{fig:comparison_smooth_LMC}, our model stream is very similar to that of \citet{2007MNRAS.380..749F}. At the leading end, the tail is also spread out wider than the part close to the cluster, which is due to approaching the apogalacticon as mentioned in \citet{2007MNRAS.380..749F}. It can be seen that the mock stream has a trend in phase-space generally in agreement with that of the data within uncertainties. What's more, S1 has a declination that the mock stream can not reach, implying it might be an outlier as well. However, S1 is exactly located at the ``deviation'' area mentioned by \citet{2006ApJ...639L..17G}, who argued that it might be caused by irregularities of the Galactic halo potential. Therefore, a smooth potential is not enough to reproduce this ``deviation'' and taking into account sub-halo component might be necessary.

\subsection{Adding the Large Magellanic Cloud}
\label{subsec:LMC}

\citet{2022MNRAS.510.2437E} has discussed the effects of dwarf galaxies on the tidal tails of Globular Clusters. They claimed that NGC 5466 stream could be perturbed due to one or more close interactions with the Large Magellanic Cloud (LMC). Motivated by this, we examine the smooth Galactic potential with an addition of LMC. This is also done through \texttt{Gala}.

\begin{figure*}
	\centering
	\includegraphics[width=\linewidth]{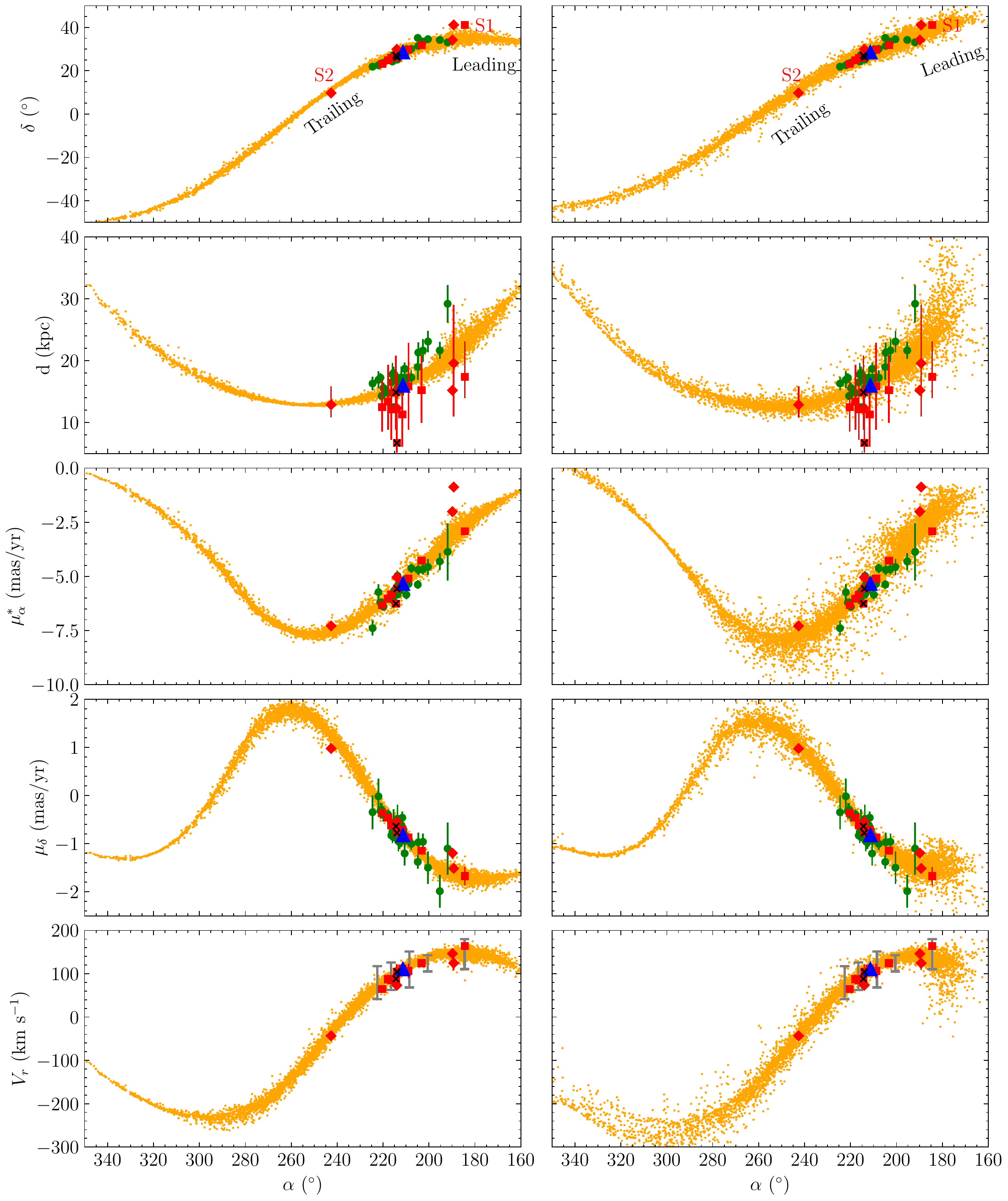}
	\caption{The planes of declination, heliocentric distance, proper motion in $\alpha$ and $\delta$, and radial velocity as a function of $\alpha$, are shown from the top to the bottom, respectively. The orange dots in the left/right panel represent the mock stream generated under a Galactic potential without/with LMC. The green points mark the members from \citet{2021MNRAS.507.1923J}. The member stars selected in Section~\ref{sec:search} are plotted in red squares. The red diamonds represent additional members identified based on the mock stream. The blue triangle represents NGC 5466 cluster. The gray error bars in the radial velocity planes correspond to $V_r^{st} \pm 3\sigma_{V_r}^{st}$.}
	\label{fig:comparison_smooth_LMC}
\end{figure*}

Following \citet{2022MNRAS.510.2437E}, we take a Hernquist Potential \citep{1990ApJ...356..359H} as the internal potential of LMC: 

\begin{equation}
    \Phi_{\rm LMC}(r') = \frac{-GM_{\rm LMC}}{r'+a_{\rm LMC}}
\end{equation}
where $r'$ is the distance to the LMC center and $M_{\rm LMC}$ and $a_{\rm LMC}$ are set to $10^{11} M_{\sun}$ and 10.2 kpc as well. The position and velocity of LMC are taken from \citet{2018A&A...616A..12G}. In the smooth potential of Section~\ref{subsec:smooth} accompanied with a moving LMC, the cluster is initialized 5 Gyr ago and then integrated forward from then on. The mock stream under this condition is displayed in the right panel of Fig.~\ref{fig:comparison_smooth_LMC}. 

As is shown, adding LMC makes the new mock stream more diffuse and a little different from the old one, especially for the end at $\alpha \sim 180\degr$. Unlike the old mock stream that bends down at the end, the counterpart of the new stream is upturned instead, reaching where S1 is. Having also examined other sub-halos such as the Small Magellanic Cloud, Draco, Ursa Minor, Fornax and Sagittarius, it is only adding LMC, the most massive satellite, that can generate a mock stream in concordance with the observation. Hence the ``deviation'' of NGC 5466 tidal stream is possibly due to the effect from LMC. 

We also estimate the pericentre ($R_{per}$), apocentre ($R_{apo}$) and eccentricity ($e$) of NGC 5466 cluster under both conditions. For the smooth potential, $R_{per} = 6.52$ kpc, $R_{apo} = 41.52$ kpc and $e = 0.73$ are very close to those estimated in \citet{2021MNRAS.507.1923J}. For the LMC-added potential, $R_{per} = 4.30$ kpc, $R_{apo} = 53.34$ kpc and $e = 0.85$ show a more elliptical orbit. We also note that the minimum separation of the cluster and LMC in the past 5 Gyr is about 22 kpc and happened 138 Myr ago, which agrees closely with that of \citet{2022MNRAS.510.2437E}. Such a close interaction also implies that the stream might have been perturbed by LMC. 

It is worth nothing that the mock stream has a longer trailing tail than the leading tail, which implies that there might be more member stars on trailing side. Moreover, the stream's track with $\alpha \ga 230\degr$ has been beyond the sky coverage of data used in both of \citet{2006ApJ...639L..17G} and \citet{2021MNRAS.507.1923J}. Therefore, we investigate this possibility using the mock stream considering its good consistency with observation. We directly restrict the data according to the mock stream's trends in phase-space. The data with $180\degr < \alpha < 260\degr$\footnote{There are no more spectroscopic data overlapping with the stream on the sky when $\alpha$ goes $> 260\degr$.} overlapping with the mock stream on the sky are retrieved and processed with CMD and metallicity cuts similar to those of Section~\ref{sec:search}. The remained stars are then selected in the plane of \{$\alpha$, $\delta$\} vs. \{$\mu_{\alpha}^*$, $\mu_{\delta}$, $V_r$\}, similar to the procedure of Fig.~\ref{fig:relation} but with the mock stream as the indicator. In this way, all members that have been selected in Section~\ref{sec:search} can be reproduced and 4 more stars are identified as shown with the red diamonds in Fig.~\ref{fig:comparison_smooth_LMC}. Among those 4 members, one is a blue horizontal branch (BHB) star whose distance can be obtained from \citet{2011ApJ...738...79X} and \citet{2017A&A...604A.108M}, and one is at $\alpha \simeq 242.65\degr$ (``S2'' hereafter) whose distance is obtained using $Gaia$ parallax = 0.0777 mas with a small relative error = 0.1868 < 20\%, and the distances of the other two stars still come from \citet{2019A&A...628A..94A}. Clearly, while eliminating the field stars, constraining sky position of the data with a narrow area created by the polynomial in Section~\ref{subsec:sky} also excludes some stream stars. 

We also note that among 9 candidate members outside the tidal radius $r_t$ obtained in Section~\ref{sec:search}, 2 stars are contaminators which have an inconsistent distance or metallicity and 4 stars have been identified in \citet{2021MNRAS.507.1923J}. Together with 4 additional members, the detailed information of 13 stars are summarized in Table~\ref{tab:members}. 

\begin{table*}
	\centering
	\caption{Identified member stars of the NGC 5466 tidal stream. Common stars of \citet{2021MNRAS.507.1923J} and this work are marked with ``$\star$''. Contaminators are marked with ``$\times$''. Additional members searched based on the mock stream are marked with ``$+$''. Heliocentric distances come from \citet{2019A&A...628A..94A} except S1, BHB and S2 (see text). The last column indicates which survey the radial velocity and metallicity come from.}
	\label{tab:members}
	\begin{tabular}{lcccccccc} % four columns, alignment for each
		\hline
		No. & $\alpha_{J2000}$ & $\delta_{J2000}$ & $d$ & $\mu_{\alpha}^*$ & $\mu_{\delta}$ & $V_r$ & [Fe/H] & Survey \\
		   & (\degr) & (\degr) & (kpc) & (mas yr$^{-1}$) & (mas yr$^{-1}$) & (km\,s$^{-1}$) & (dex) &  \\
		\hline
		1 (S1) & 184.4038 & 41.1251 & $17.37^{+5.74}_{-3.45}$ & $-2.9170\pm 0.1490$ & $-1.6745\pm 0.1860$ & $163.91\pm 6.19$ & $-1.943 \pm 0.044$ & SEGUE \\
		2$^+$ & 189.2492 & 41.1226 & $19.58^{+9.48}_{-8.58}$ & $-0.8765\pm 0.0543$ & $-1.5142\pm 0.0516$ & $124.60\pm 15.65$ & $-1.968\pm 0.264$ & LAMOST \\
		3$^+$ (BHB) & 189.8104 & 34.1988 & $15.20^{+0.40}_{-0.40}$ & $-2.0117\pm 0.0588$ & $-1.1971\pm 0.0529$ & $146.42\pm 2.58$ & $-1.995\pm 0.055$ & SEGUE \\
		4 & 203.2958 & 31.8642 & $15.23^{+5.67}_{-5.28}$ & $-4.2747\pm 0.0180$ & $-1.1452\pm 0.0157$ & $124.58\pm 9.48$ & $-2.085\pm 0.149$ & LAMOST \\
		5$^\star$ & 208.9050 & 29.6897 & $15.82^{+7.09}_{-6.96}$ & $-5.1124\pm 0.0710$ & $-0.8719\pm 0.0519$ & $106.30\pm 16.71$ & $-2.046\pm 0.356$ & LAMOST \\
		6$^\star$ & 211.6564 & 28.2157 & $11.29^{+4.34}_{-5.20}$ &$-5.2769\pm 0.0407$ & $-0.7132\pm 0.0334$ & $111.78\pm 9.06$ & $-1.901\pm 0.105$ & LAMOST \\
		7$^\times$ & 213.9924 & 26.1863 & $6.69^{+2.16}_{-1.60}$ & $-5.5692\pm 0.0441$ & $-0.7774\pm 0.0480$ & $104.40\pm 17.56$ & $-1.602\pm 0.096$ & LAMOST \\
		8$^+$ & 214.0078 & 29.8912 & $12.31^{+5.55}_{-3.73}$ & $-5.0479\pm 0.0281$ & $-0.5467\pm 0.0248$ & $73.43\pm 8.99$ & $-2.018\pm 0.127$ & LAMOST \\
		9$^\times$ & 214.4116 & 27.0612 & $14.86^{+6.00}_{-6.04}$ & $-6.2586\pm 0.0500$ & $-0.6391\pm 0.0449$ & $88.35\pm 10.33$ & $-1.542\pm 0.120$ & LAMOST \\
		10$^\star$ & 216.4128 & 25.8587 & $12.10^{+4.98}_{-4.91}$ & $-5.8371\pm 0.0324$ & $-0.6180\pm 0.0310$ & $84.85\pm 7.35$ & $-1.876\pm 0.023$ & LAMOST \\
		11$^\star$ & 217.8405 & 24.8631 & $13.40^{+5.92}_{-4.54}$ &$-6.0359\pm 0.0369$ & $-0.4615\pm 0.0395$ & $87.31\pm 10.33$ & $-1.954\pm 0.125$ & LAMOST \\
		12 & 220.3912 & 23.1246 & $12.47^{+4.35}_{-3.93}$ & $-6.3277\pm 0.0212$ & $-0.3645\pm 0.0270$ & $64.90\pm 7.50$ & $-1.981\pm 0.023$ & LAMOST \\
		13$^+$ (S2) & 242.6515 & 9.7092 & $12.87^{+2.96}_{-2.03}$ & $-7.2878\pm 0.0131$ & $0.9768\pm 0.0126$ & $-43.32\pm 5.71$ & $-1.929\pm 0.032$ & LAMOST \\
		\hline
	\end{tabular}
\end{table*}

\begin{figure}
	\includegraphics[width=\columnwidth]{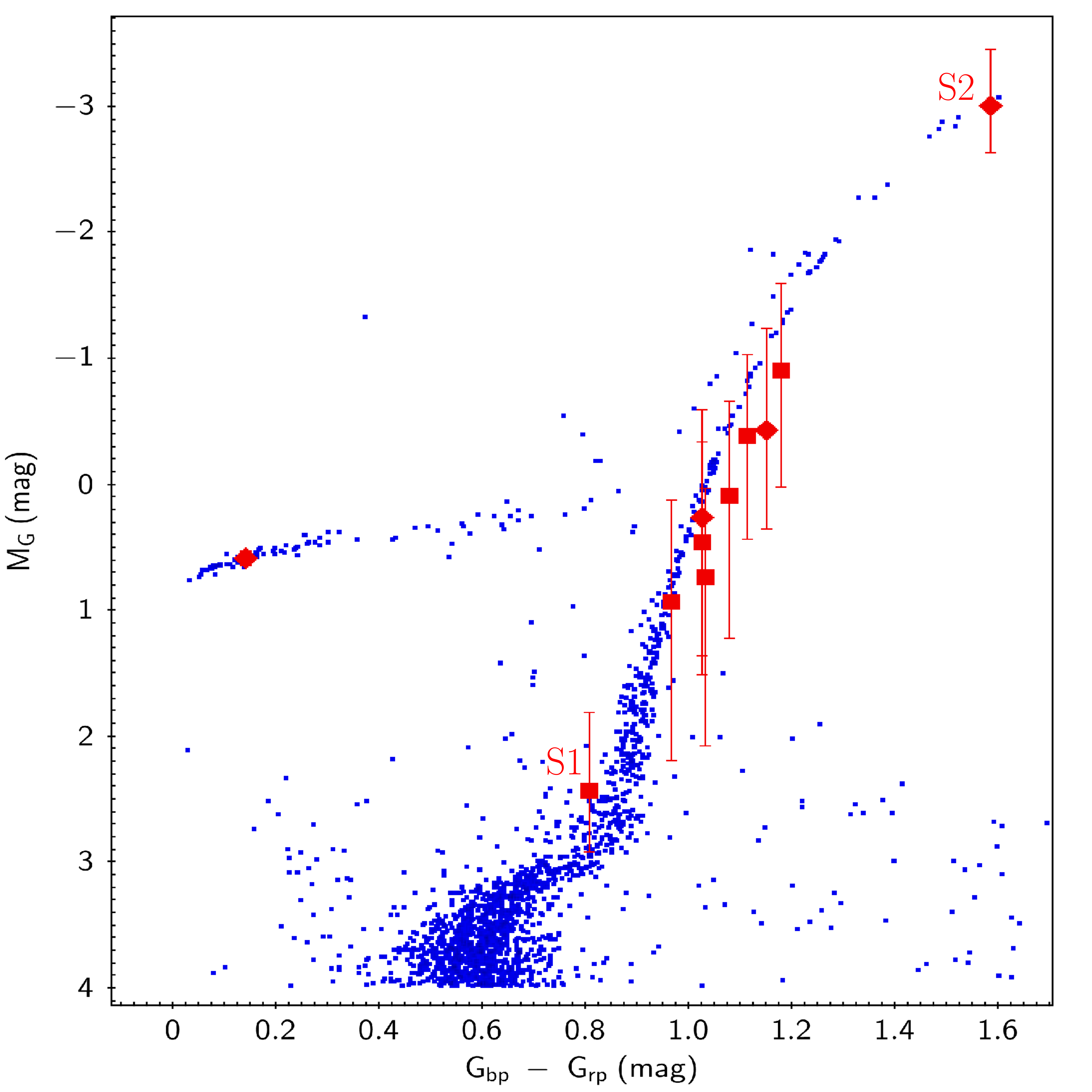}
	\caption{The blue dots represent the stars within the tidal radius of NGC 5466. The member stars are plotted in the red squares and diamonds which have the same meaning as Fig.~\ref{fig:comparison_smooth_LMC}. The error bars account for uncertainties in absolute magnitude $M_G$ due to the distance errors. The two contaminators are not plotted here.}
	\label{fig:CMD_done}
\end{figure}

Consistent trends in phase-space between the model stream and our members have been shown in Fig.~\ref{fig:comparison_smooth_LMC} and a metallicity of [Fe/H] = -1.98 dex of the cluster is also in concordance with [Fe/H] of stars in Table~\ref{tab:members}. Another inspection may be about stellar age which can be reflected by the $Gaia$ photometry. Fig.~\ref{fig:CMD_done} shows the color - absolute magnitude diagram of the cluster and our stars, where the error bars account for uncertainties in $M_G$ due to the distance errors. All member stars including S1 (darkest) and S2 (brightest) lie in the cluster's isochrone well within uncertainties. 

\section{Discussion}
\label{sec:discussion}

As introduced in Section~\ref{sec:introduction}, different results on the length of this tenuous stream have been concluded. To break the deadlock, this work is dedicated to search for the stream's member stars which could be a direct indicator of the stream's length. Compared to a $31\degr$ tail traced by \citet{2021MNRAS.507.1923J}, our results contend a $\sim 60\degr$ long stream from S1 at leading side to S2 at trailing side. S1 is $\sim 27\degr$ away from the cluster ($\alpha = 211.3640\degr$), which supports the statement of \citet{2006ApJ...639L..17G} that the stream extends $\sim 30\degr$ towards the leading direction. S2 has a celestial position of $(\alpha, \delta)$ = $(242.6515\degr, 9.7092\degr)$ beyond the sky coverage of data used in \citet{2006ApJ...639L..17G} and \citet{2021MNRAS.507.1923J}.

\begin{figure*}
	\includegraphics[width=0.85\linewidth]{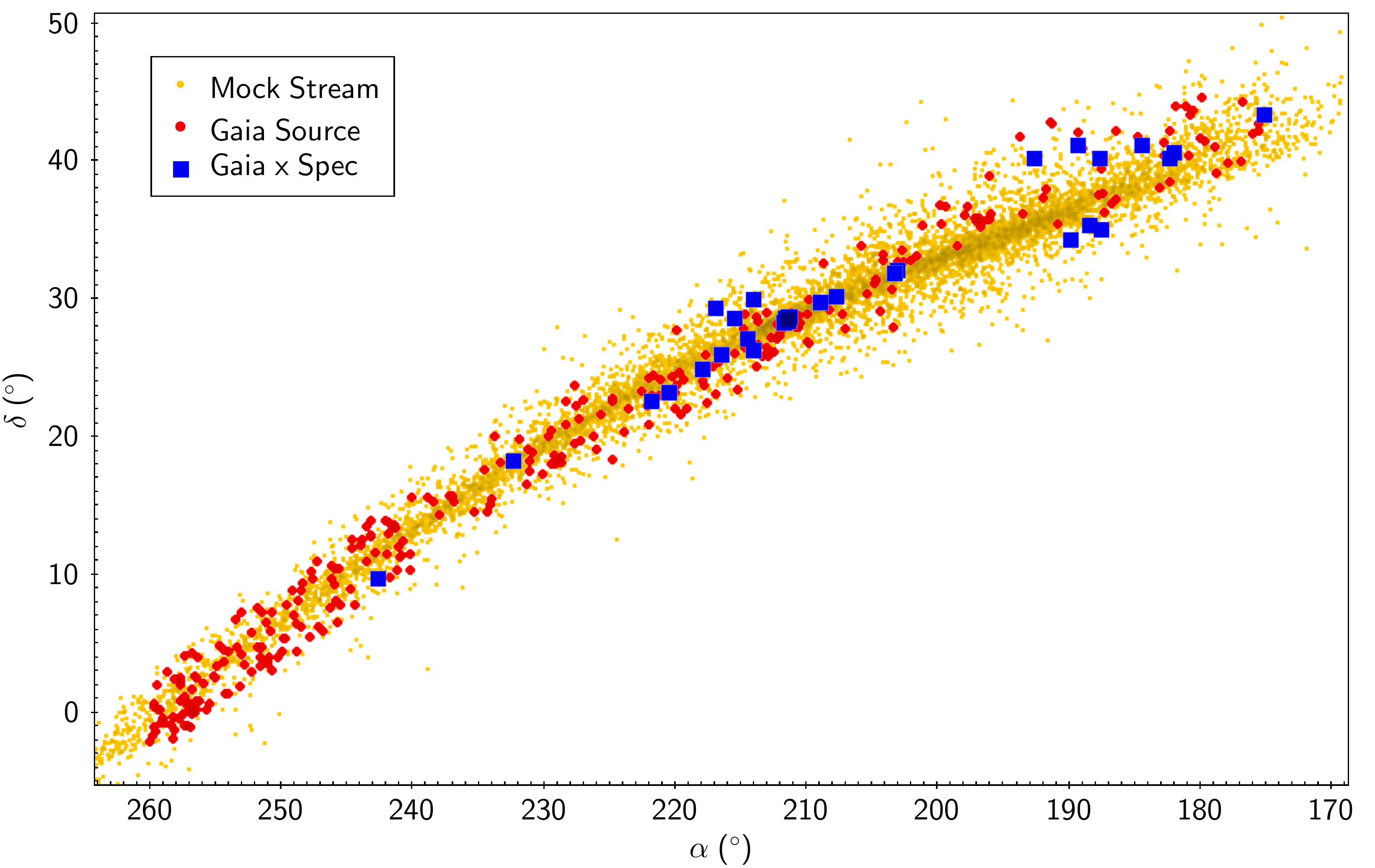}
	\caption{The orange points represent the mock stream. The red points represent the stream candidates which are obtained with sky position, CMD and proper motion cuts based on the model stream just using $Gaia$ data. The blue squares represent common stars between $Gaia$ and LAMOST+SEGUE data.}
	\label{fig:GaiaxSpec}
\end{figure*}

Although a longer stream is traced in this work, it should be noted that some of members are far apart from the others. For example, S2 and No.12 are spaced by 22$\degr$ in $\alpha$, between which there is no star detected. We consider that this might be caused by incompleteness of the spectroscopic data, and we do some tests to illustrate it. In Fig.~\ref{fig:GaiaxSpec}, just using $Gaia$ data, we obtain about 850 stream candidates through sky position, CMD and proper motion cuts based on the model stream and plot them with the red points. It can be seen that those candidates are generally continuously distributed in space, although there must be contaminants because they still need to be selected through metallicity and radial velocity which are not provided by $Gaia$. The number of stars is reduced substantially after cross-matching with LAMOST and SEGUE data, only 36 common stars remained as shown with the blue squares in the figure. There are large gaps now. Therefore, we think the number of spectra is very limited and most (even more than 95\%) of $Gaia$ sources are lost during cross-matching, which probably include stream stars that we need. In other words, some stream stars are not observed in those spectroscopic surveys. However, we can not completely exclude that the stream truly has large spatial gaps due to some physical mechanism, because the data about the stream we have are limited yet. It is expected that this problem could be solved through more detailed observations provided by the upcoming surveys. 

Individual stream members can be used to constrain the Galactic potential \citep[e.g.,][]{2013MNRAS.436.2386L,2015ApJ...803...80K,2016ApJ...833...31B,2019MNRAS.486.2995M}. Therefore, the search for those members are fundamental and crucial. We anticipate that our results could be helpful in detecting more stream stars of NGC 5466 in the later observations, such as the follow-up spectroscopic campaign PFS-SSP (Subaru Strategic Program with Prime Focus Spectrograph) with deeper limiting magnitude.

\section{Summary}
\label{sec:summary}

With astrometric and photometric data from $Gaia$ EDR3, spectroscopic data from SDSS/SEGUE DR16 and LAMOST DR8, 11 member stars of the NGC 5466 tidal stream are identified after applying a variety of cuts in sky position, color-magnitude diagram, metallicity, proper motion and radial velocity. Among the 11 stars, 4 have been detected by \citet{2021MNRAS.507.1923J} and 7 are newly identified. By modeling the cluster's disruption, a mock stream in concordance with observation is reproduced under a smooth Galactic potential combined with the Large Magellanic Cloud, which might have perturbed the stream. The detection of S1 at leading side supports the claim of \citet{2006ApJ...639L..17G}. From S1 to S2, another highly probable member at trailing tail, the stream's length is traced to $60\degr$ of sky. Given that NGC 5466 is so distant and potentially has a longer tail than previously thought, we expect that like Pal 5 and GD-1, NGC 5466 tidal stream will be another useful tool in constraining the Milky Way potential.

\section*{Acknowledgements}

We thank the referee for the thorough reviews that have helped us to improve the manuscript. This study is supported by the National Natural Science Foundation of China under grant No. 11988101, 11973048, 11927804, 11890694 and National Key R\&D Program of China No. 2019YFA0405502. We acknowledge the support from the 2m Chinese Space Station Telescope project: CMS-CSST-2021-A10, CMS-CSST-2021-B05. 

Guoshoujing Telescope (the Large Sky Area Multi-Object Fiber Spectroscopic Telescope LAMOST) is a National Major Scientific Project built by the Chinese Academy of Sciences. Funding for the project has been provided by the National Development and Reform Commission. LAMOST is operated and managed by the National Astronomical Observatories, Chinese Academy of Sciences.

This work presents results from the European Space Agency (ESA) space mission $Gaia$. $Gaia$ data are being processed by the $Gaia$ Data Processing and Analysis Consortium (DPAC). Funding for the DPAC is provided by national institutions, in particular the institutions participating in the $Gaia$ MultiLateral Agreement (MLA). The $Gaia$ mission website is \url{https://www.cosmos.esa.int/gaia}. The $Gaia$ archive website is \url{https://archives.esac.esa.int/gaia}.

Funding for the Sloan Digital Sky Survey IV has been provided by the Alfred P. Sloan Foundation, the U.S. Department of Energy Office of Science, and the Participating Institutions. 

SDSS-IV acknowledges support and resources from the Center for High Performance Computing  at the University of Utah. The SDSS website is \url{www.sdss.org}.

SDSS-IV is managed by the Astrophysical Research Consortium for the Participating Institutions of the SDSS Collaboration including the Brazilian Participation Group, the Carnegie Institution for Science, Carnegie Mellon University, Center for Astrophysics | Harvard \& Smithsonian, the Chilean Participation Group, the French Participation Group, Instituto de Astrof\'isica de Canarias, The Johns Hopkins University, Kavli Institute for the Physics and Mathematics of the Universe (IPMU) / University of Tokyo, the Korean Participation Group, Lawrence Berkeley National Laboratory, Leibniz Institut f\"ur Astrophysik Potsdam (AIP),  Max-Planck-Institut f\"ur Astronomie (MPIA Heidelberg), Max-Planck-Institut f\"ur Astrophysik (MPA Garching), Max-Planck-Institut f\"ur Extraterrestrische Physik (MPE), National Astronomical Observatories of China, New Mexico State University, New York University, University of Notre Dame, Observat\'ario Nacional / MCTI, The Ohio State University, Pennsylvania State University, Shanghai Astronomical Observatory, United Kingdom Participation Group, Universidad Nacional Aut\'onoma de M\'exico, University of Arizona, University of Colorado Boulder, University of Oxford, University of Portsmouth, University of Utah, University of Virginia, University of Washington, University of Wisconsin, Vanderbilt University, and Yale University.

%%%%%%%%%%%%%%%%%%%%%%%%%%%%%%%%%%%%%%%%%%%%%%%%%%
\section*{Data Availability}

The identified member stars of the NGC 5466 tidal stream are available in the article.



\bibliographystyle{mnras}
\bibliography{revised_manuscript} % if your bibtex file is called example.bib

%\begin{thebibliography}{99}
%\end{thebibliography}

%%%%%%%%%%%%%%%%%%%%%%%%%%%%%%%%%%%%%%%%%%%%%%%%%%

%%%%%%%%%%%%%%%%% APPENDICES %%%%%%%%%%%%%%%%%%%%%

%\appendix



%%%%%%%%%%%%%%%%%%%%%%%%%%%%%%%%%%%%%%%%%%%%%%%%%%

% Don't change these lines
\bsp	% typesetting comment
\label{lastpage}
\end{document}